\documentstyle[epsfig]{mn}

\begin{document}

\title[Cygnus X-3 with {\rm ISO}]
{Cygnus X-3 with \bf{\it{ISO\/}}: investigating the wind}  
\author[R. N. Ogley, S. J. Bell Burnell and R. P. Fender]
{R. N. Ogley$^{1,2}$, 
S. J. Bell Burnell$^1$ and 
R. P. Fender$^3$\\
$^1$Department of Physics and Astronomy, The Open University, Milton
Keynes, MK7 6AA, UK.\\
$^2$Present address: Service d'Astrophysique, CEA Saclay, Orme des Merisiers - B\^{a}t 709, F-91191 Gif sur Yvette, Cedex, France.\\
$^3$Astronomical Institute `Anton Pannekoek' and Center for High Energy
Astrophysics, University of Amsterdam, Kruislaan 403,\\
1098 SJ Amsterdam, The Netherlands.}

\maketitle
\begin{abstract}
We observed the energetic binary Cygnus X-3 in both quiescent and
flaring states between 4 and 16 $\mu$m using the {\it ISO} satellite.
We find that the quiescent source shows the thermal free-free spectrum
typical of a hot, fast stellar wind, such as from a massive helium
star.  The quiescent mass-loss rate due to a spherically symmetric,
non-accelerating wind is found to be in the range (0.4--2.9)$\times
10^{-4}$ $\rm M_{\odot}$ yr$^{-1}$, consistent with other infrared and
radio observations, but considerably larger than the $10^{-5}$ $\rm
M_{\odot}$ yr$^{-1}$ deduced from both the orbital change and the
X-ray column density.  There is rapid, large amplitude flaring at 4.5
and 11.5 $\mu$m at the same time as enhanced radio and X-ray activity,
with the infrared spectrum apparently becoming flatter in the flaring
state.  We believe non-thermal processes are operating, perhaps along
with enhanced thermal emission.

\end{abstract}

\begin{keywords}
stars: individual: CygX-3 -- stars: Wolf-Rayet -- binaries: close --
stars: mass-loss -- infrared: stars -- X-rays: stars
\end{keywords}

\section{Introduction}

The Infrared Space Observatory {\it ISO} (Kessler et al.\ 1996) with
its on-board complement of cameras and spectrometers, provided a rare
opportunity to study sources at wavelengths not accessible from the
ground. We present photometry and images of the Cygnus X-3 field made
using ISOCAM (Cesarsky et al.\ 1996), the camera on {\it ISO}, in the
wavelength range 4 to 16 $\mu$m.  We report the first spectrum of this
source in this wavelength range.

Cygnus X-3 is an enigmatic and unusual X-ray binary which was
discovered $>$30 years ago (Giacconi et al.\ 1967). Observations of it
have been made from radio to gamma rays and in every wavelength bands
the system exhibits unusual behaviour.  In X-rays and infrared, the
emission is partially modulated by a 4.8 h period (Mason, C\'{o}rdova
\& White 1986), and this is believed to be the orbital
period of the system.

In 1991, van Kerkwijk et al. observed Cyg X-3 in the infrared with the
CGS4 spectrometer on UKIRT and discovered He\,{\sc i} and He\,{\sc ii}
together with N emission lines which were modulated at the 4.8 h
period (van Kerkwijk et al.\ 1992; van Kerkwijk et al.\ 1996).  They
concluded that the lines originated from the companion star, either a
Wolf--Rayet of the WN7 class, or an earlier WN4--6 in an outburst
phase.  This was investigated in more detail by Fender, Hanson \&
Pooley (1999, hereafter FHP99), who took spectra during a variety of
activity states and confirmed that the underlying quiescent spectral
type is of an earlier WN4--5 Wolf--Rayet star.

Galactic Wolf--Rayet stars have high velocity winds, up to 2500 km
s$^{-1}$ (Crowther, Hillier \& Smith 1995; Bohannan \& Crowther 1999).
The Wolf--Rayet wind from Cyg X-3 is believed to be the main
contributor in providing mass and angular momentum loss which increase
the orbital period ($\dot{P} = -6.6 \times 10^{-10}$ s s$^{-1}$, Kitamoto et al.\ 1995).  This period increase is consistent with a
mass-loss rate of $10^{-6} M_{\rm T}$ $\rm M_{\odot}$ yr $^{-1}$,
where $M_{\rm T}$ is the total mass in the system (Kitamoto et al.\
1995; Matz 1997).  Kitamoto et al.\ (1994) modelled changes in the
radio activity, infrared and X-ray states caused by changes in the
density of the WR wind (Table \ref{wind}).  The multi-wavelength
predictions of Kitamoto et al.\ (1994) were tested by FHP99, who find
that during an outburst the infrared {\it K} band spectrum becomes
dominated by strong twin-peaked \hbox{He\,{\sc i}} emission lines.
These lines show strong {\it V}/{\it R} variability correlated with
the orbital phase.  Due to the small size of the binary ($\sim
5\;R_{\odot}$ separation, 4.8 hour period) the observed line flux could
not arise from a region of this size as the environment would
be too hot.  FHP99 conclude that the He\,{\sc i} emission lines are
emitted from a region which is significantly larger than the binary
separation, and that the variability of these lines indicate an
asymmetric emission region, possibly from a flattened disc-like wind.

\begin{table}
\caption{Radio, infrared and X-ray variability based on a changing
density wind, from Kitamoto et al.\ (1994).  The term `modulation'
refers to the continuum.  The states marked with an asterisk (*) have
been expanded on and confirmed by Fender et al.\ (FHP99).}
\begin{tabular}{llll}\hline
Regime 		& Emission 	& Rare wind 	& Dense wind \\ \hline
Radio		& Continuum*	& Quiescence	& Flaring \\
Infrared	& Continuum*	& Weak		& Strong \\
		& Line strength*& Weak		& Strong \\
		& Ionisation*	& High		& Low	\\
		& Modulation	& Large		& Small	\\
X-ray		& Luminosity*	& Low		& High	\\
		& Spectrum	& Hard		& Soft 	\\ \hline
\end{tabular}
\label{wind}
\end{table}

The compact object in Cyg X-3 is less well defined.  Cherepashchuk \&
Moffat (1994) suggest that the high X-ray luminosity (greater than the
Eddington luminosity for a 1 ${\rm M_{\odot}}$ object) implies that
the compact object is a black hole.  Schmutz, Geballe \& Schild (1996)
related time-variations of infrared spectral lines to an orbital
velocity and derived a mass for the compact object in the range of
7--40 $\rm M_{\sun}$, assuming realistic values for the mass of the
Wolf--Rayet star and the inclination of the system.  However, their
model does not account for all the relative phasing of the X-rays and
infrared lines.  Ergma \& Yungelson (1998) point out that if the
progenitor to the binary is two massive helium stars then the
mass-loss and common envelope phase of these two stars prevents the
formation of a black hole \& Wolf--Rayet binary with an orbital period
less than several days.  This ambiguity does not resolve itself if the
compact object is a neutron star; Ergma \& Yungelson (1998) also
comment that the propeller motion of a spinning neutron star would
prevent wind accretion in the system, thus removing the jet-creation
mechanism.

With a high mass, high mass-loss (wind) object in close proximity to a
compact object, a high degree of interaction takes place.  Radio
observations at wavelengths ranging from 1--20 GHz have shown Cyg X-3
to undergo many states of emission and are used to define the state of
Cyg X-3.  For most of its time, the system occupies a {\it quiescent}
radio state, with 8.3 GHz flux around 100 mJy (Waltman et al.\ 1994).
{\it Major flares} (8.3 GHz flux $>$ 1 Jy) occur at random intervals
every 18 months or so and relaxation after a flare can last up to
several weeks (Waltman et al.\ 1995).  More common are {\it minor
flares} (8.3 GHz flux 200--1000 mJy) and these usually occur during a
flaring period typically several weeks duration, with several small
flares occurring in close proximity to each other (Waltman et al.\
1996).  Minor flares can reach peak emission in a few hours and decay
over several hours (Ogley 1998).  The minor and major flares have been
shown to be intrinsically different in their production, rise and
relaxation times (Newell 1996).  Preceding a major flare, the system
enters a quenching state with fluxes at 8.3 GHz dropping to around
10 mJy.  The system returns to quiescence then flares (Waltman et al.\
1996).

The main mechanism for radio flares is thought to be associated with
the acceleration of relativistic electrons in the base of jets.
Synchrotron emitting plasmons have been detected, travelling at
$\simeq$ 5 mas d$^{-1}$ on both arcsec and mas scales (Spencer et
al.\ 1986; Mioduszewski et al.\ 1998).  There is no confident
determination of the velocity of ejecta: while a velocity of 0.3$c$ is
generally taken as the speed on the larger arcsec scale (e.g. Spencer et
al.\ 1986), other highly-relativistic interpretations are possible on
the mas scale (e.g. Ogley, Bell Burnell \& Newell 1997; Newell, Garrett
\& Spencer 1998).

The Wolf--Rayet contributes not only to the enigmatic qualities of the
system, but also a significant and varying opacity.  The true nature
of the secondary is a top priority for the understanding of the
system.  Observations at the {\it ISO} wavelengths provide a unique
opportunity to study the Wolf--Rayet wind, and spectra obtained can
track its true nature.

\section{ISOCAM observations}

\subsection{In quiescence}

Quiescent observations were taken on 1996 April 07 (MJD 50180) with 6
long-wavelength (LW) filters on the CAM instrument.  Cyg X-3 is in the
galactic plane at a distance of around 10-kpc and so is heavily
obscured by dust along our line of sight.  Around the object, emission
from cirrus is particularly strong at longer wavelengths so our
observations were focused in the 1--16 $\mu$m range.  The wavelengths
chosen are presented in Table \ref{iso_filters}.  We observed for a
total of 2895 s in 6 filters in order to obtain a rough spectrum of
the source.  A 6 arcsec pixel size was used for the observations and
single frames, stare-mode, were taken with the 4 shorter wavelength
filters, whereas the 12 and 15 $\mu$m images were taken as $3\times 3$
mosaics.  The integration time for individual frames was 2.1 s and the
total integration times for all filters are given in
Table~\ref{iso_filters}

\begin{table}
\caption{ISOCAM filters and integration times used in the quiescent
 observations.}
\begin{tabular}{lll}\hline
Filter 	& Wavelengths 	& Integration 	\\
name	& range ($\mu$m)& time (s)	\\
\hline
LW 1	& 4.10--4.90	& 161	\\
LW 4	& 5.50--6.50	& 161	\\
LW 5	& 6.50--7.00	& 119	\\
LW 6	& 7.00--8.50	& 144	\\
LW 10	& 8.60--14.4	& 321	\\
LW 3	& 12.1--16.9	& 311	\\
\hline
\end{tabular}
\label{iso_filters}
\end{table}

Analysis of the {\it ISO} data was performed at the {\it ISO} data
centre in Vilspa, Spain.  Each image had a number of corrections
applied to it and these included: (a) dark correction, (b)
de-glitching the data to remove transients, (c) stabilisation of
pixels due to memory effects, (d) a flat-field correction, (e) jitter
correction, (f) combination of images into a stare image or
raster map, (g) field of view distortion and finally (h) conversion
from engineering units of adu g$^{-1}$ s$^{-1}$ to Jy.

The corrections that were applied to the data either used standard
library frames, or software routines developed by the ISOCAM
Consortium.  The processes which used library fields were the dark
correction (van Buren \& Kong 1996) and the flat-field correction of
stare images.  The other corrections used standard software routines
and included a multi-resolution median transform for the de-glitching
process; a 90 per cent stabilisation method to correct for memory
effects; standard field of view distortion; and standard conversion of
adu g$^{-1}$ s$^{-1}$ to Jy with code formulated on 1998 April 05.  No
jitter correction as the average jitter reported of 0.5 arcsec was
much less than our pixel field of view of 6 arcsec.  For a detailed
discussion see Ogley (1998).


\begin{figure*}
\epsfig{file=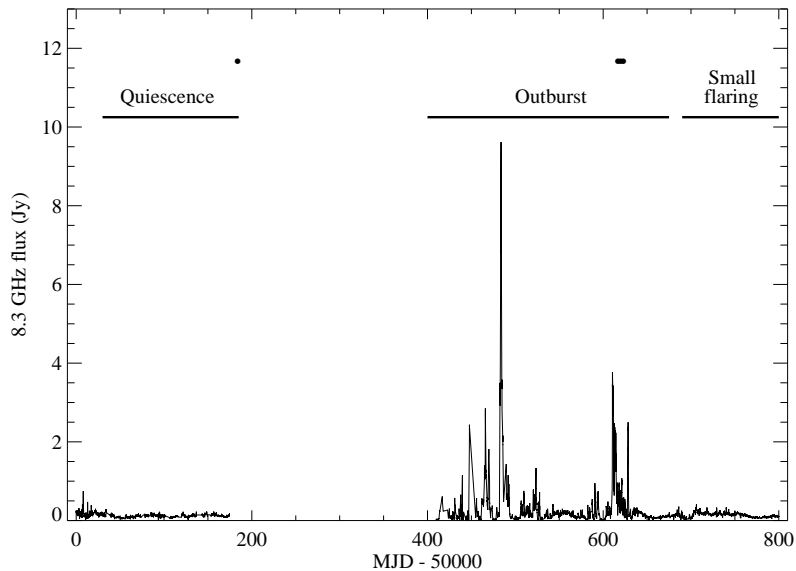, width=4.5in}
\caption{Radio photometry at 8.3 GHz from the Green Bank
Interferometer.  Displayed by the bullet points are the times of our
four ISO observations, one epoch during quiescence, and three epochs
during outburst.  No data was available between MJD 50200 and 50400 as
the GBI was off-line for financial reasons.}
\label{GBI_phot}
\end{figure*}

The reduction of mosaic images was essentially identical to stare
images, but with the following exceptions.  The flat-field correction
did not use standard frames, but built up a flat-field using the
background pixels in the mosaic.  The flat-fielding using standard
frames is the least robust of the data analysis stages, and can lead
to significant increases or decreases in flux at the edges of an
image.  Memory effects due to pixel stabilisation can create a
cross-shaped effect on mosaic images.  This was reduced by
sequentially masking the residual pixels as the mosaic is built up.

\subsection{In outburst}

Three epochs of {\it ISO} data were acquired to observe Cyg X-3 during
a major flare on 1997 Jun 15, 18, 21; MJD 50613.7, 50615.4, 50619.9.
Fig.\ \ref{GBI_phot} shows photometry at 8.3 GHz from the Green Bank
Interferometer (GBI) with the times of our {\it ISO} observations in
both quiescent and flaring states shown by bullet points, one epoch
during quiescence, and three epochs during outburst.  Each flaring
epoch consisted of a stare observation with CAM using the LW1 filter
(4.5 $\mu$m), a mosaic observation with CAM using the LW10 filter
(11.5 $\mu$m) and a PHOT observation.  Due to instrumental
difficulties only the CAM data is presented.  The procedure used in
reducing the data was identical to the quiescent epoch.

\section{Results}

\subsection{The Images}

\subsubsection{Digitized sky survey}

\begin{figure}
\epsfig{file=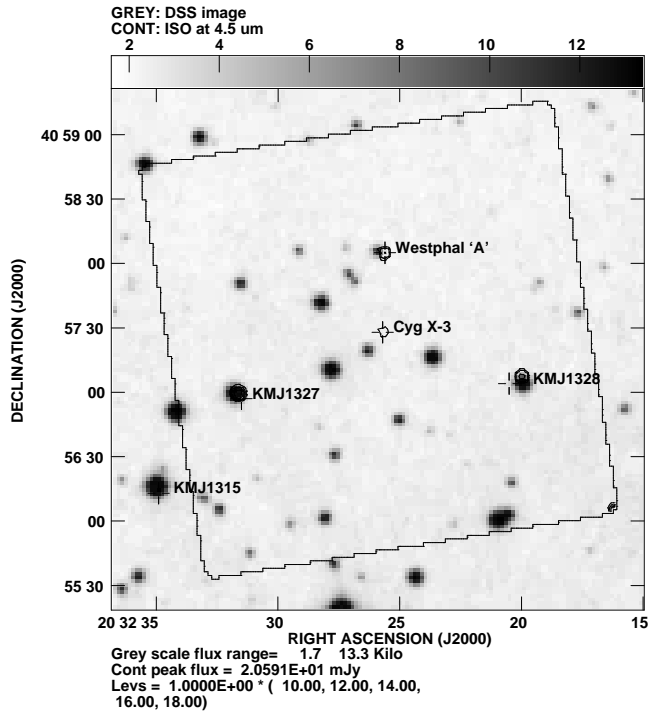, width=3.4in}
\caption{Digitized sky survey image of the Cyg X-3 area.  The box
shows the orientation and position of the stare-mode ISOCAM images.
For further details see the text.}
\label{dss}
\end{figure}

There is a striking difference between the {\it ISO} images and an
optical digitized sky survey (DSS) image of the same region.  The DSS
image, Fig.\ \ref{dss}, shows many more stars than are visible on the
{\it ISO} images (Fig.\ \ref{cam_images}).  Cyg X-3 is undetected on
the DSS plate as it has $A_{V}\sim 30$ mag of extinction, and an
estimated {\it V} band magnitude of $V = 29 \pm 1$ (Wagner et al.\
1990).  The radio position of the source is represented by a cross, at
the centre of the image.  Stars marked KMJ 1315, 1327, 1328
(Kobulnicky, Molnar \& Jones 1994) and star Westphal `A' (Westphal et
al.\ 1972) are all identified.  KMJ 1328 is Westphal's star `C' with
{\it V} = 15.03.  One can see that, as is expected along the galactic
plane, there is an abundance of sources at various magnitudes.

\begin{table*}
\caption{Photometric fluxes of all four sources, during a Cyg X-3
quiescent state.  A 12-arcsec aperture was used, corrected for the
different point-spread functions of the filters.  No data have been
de-reddened.}
\begin{tabular}{lcccccc}\hline
Star & 4.5 $\mu$m flux & 6.0 $\mu$m flux & 6.75 $\mu$m flux &
    7.75 $\mu$m flux & 11.5 $\mu$m flux & 14.5 $\mu$m flux \\ & (mJy)
    & (mJy) & (mJy) & (mJy) & (mJy) & (mJy) \\ \hline

KMJ 1327 & 35.4 $\pm$ 2.2 & 50.2 $\pm$ 3.6 & 30.9 $\pm$ 5.1 & 30.6
$\pm$ 4.9 & 8.9 $\pm$ 3.0 & 6.5 $\pm$ 2.7 \\
KMJ 1328 & 26.0 $\pm$ 1.8 & 28.0 $\pm$ 3.6 & 22.0 $\pm$ 5.7 & 17.6
$\pm$ 6.8 & 8.0 $\pm$ 2.6 & 3.9 $\pm$ 2.0 \\
Westphal et al. `A'& 21.0 $\pm$ 2.2 & 22.2 $\pm$ 3.2 & 14.2 $\pm$ 4.9
& 22.4 $\pm$ 4.9 & 7.0 $\pm$ 1.6 & 4.3 $\pm$ 1.5 \\
Cyg X-3 & 18.2 $\pm$ 2.5 & 31.1 $\pm$ 4.1 & 20.6 $\pm$ 4.9 & 24.3
$\pm$ 3.9 & 15.2$\pm$ 1.6 & 18.7$\pm$ 1.8 \\
\hline
\end{tabular}
\label{fluxes}
\end{table*}

\subsubsection{ISOCAM--quiescence}

\begin{figure*}
\vspace{-3mm}
\centerline{\epsfig{file=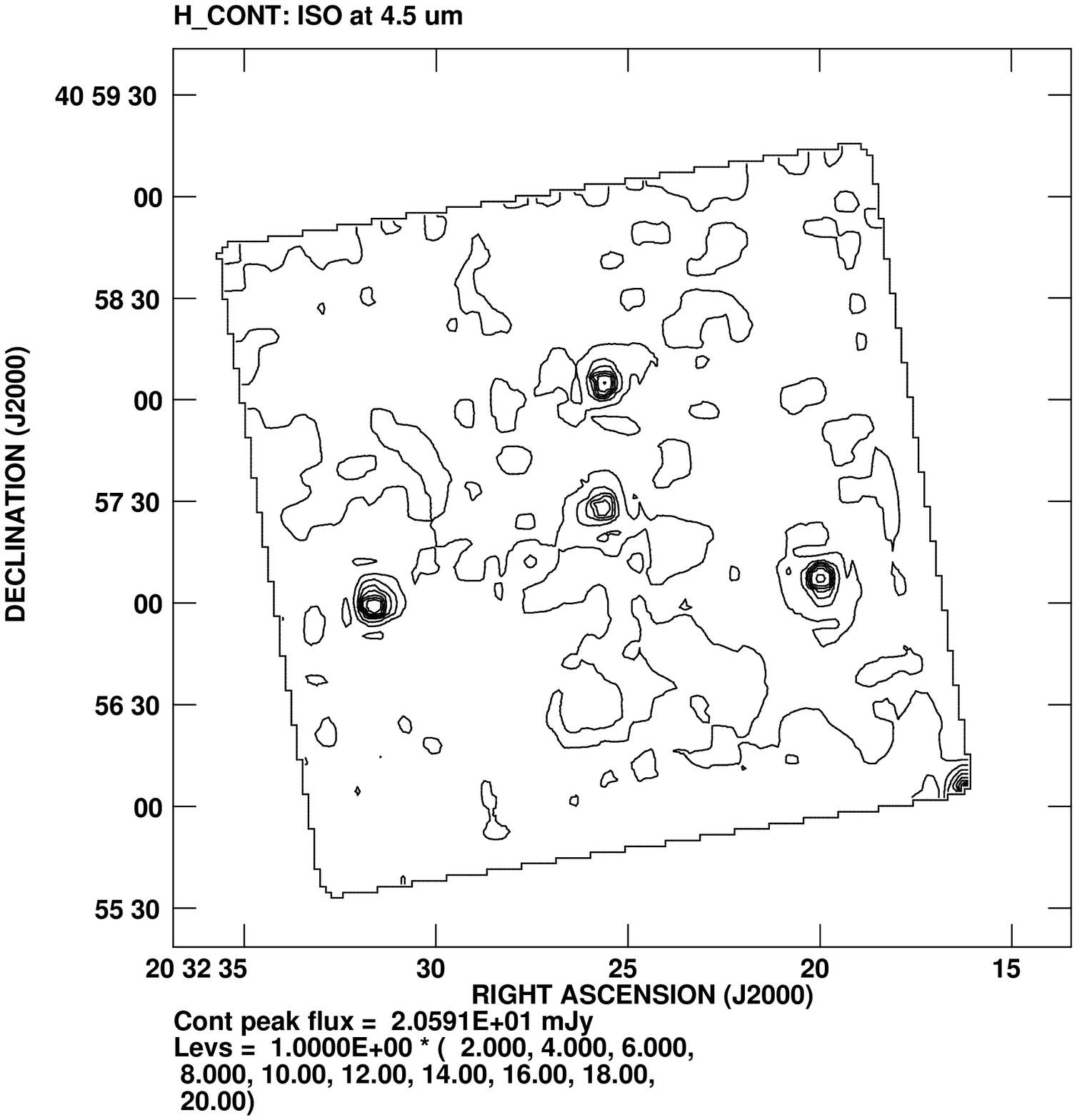,angle=0,width=76mm,clip=}\quad\epsfig{
file=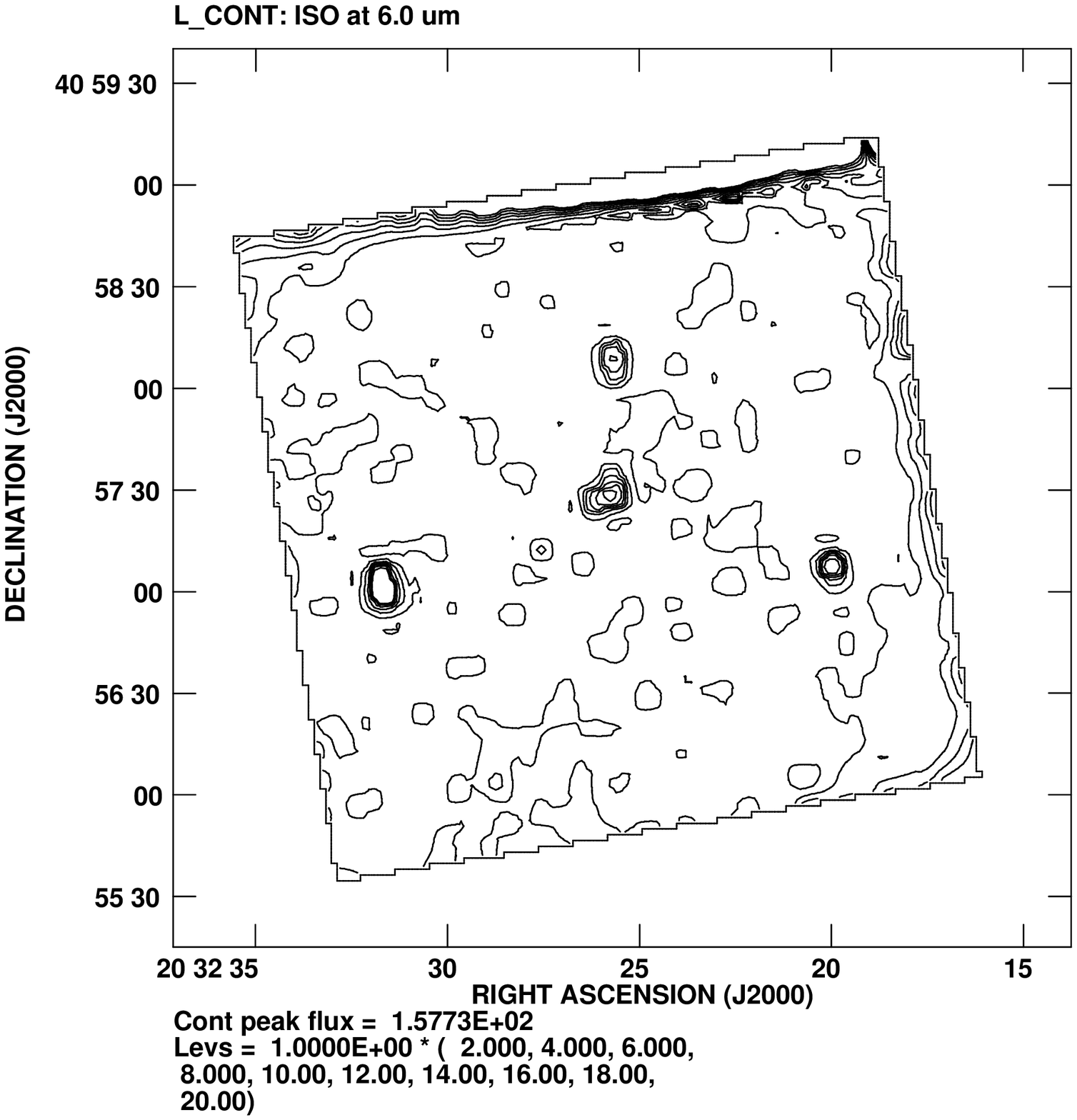,angle=0,width=76mm,clip=}}
\vspace{-1mm}
\centerline{\epsfig{file=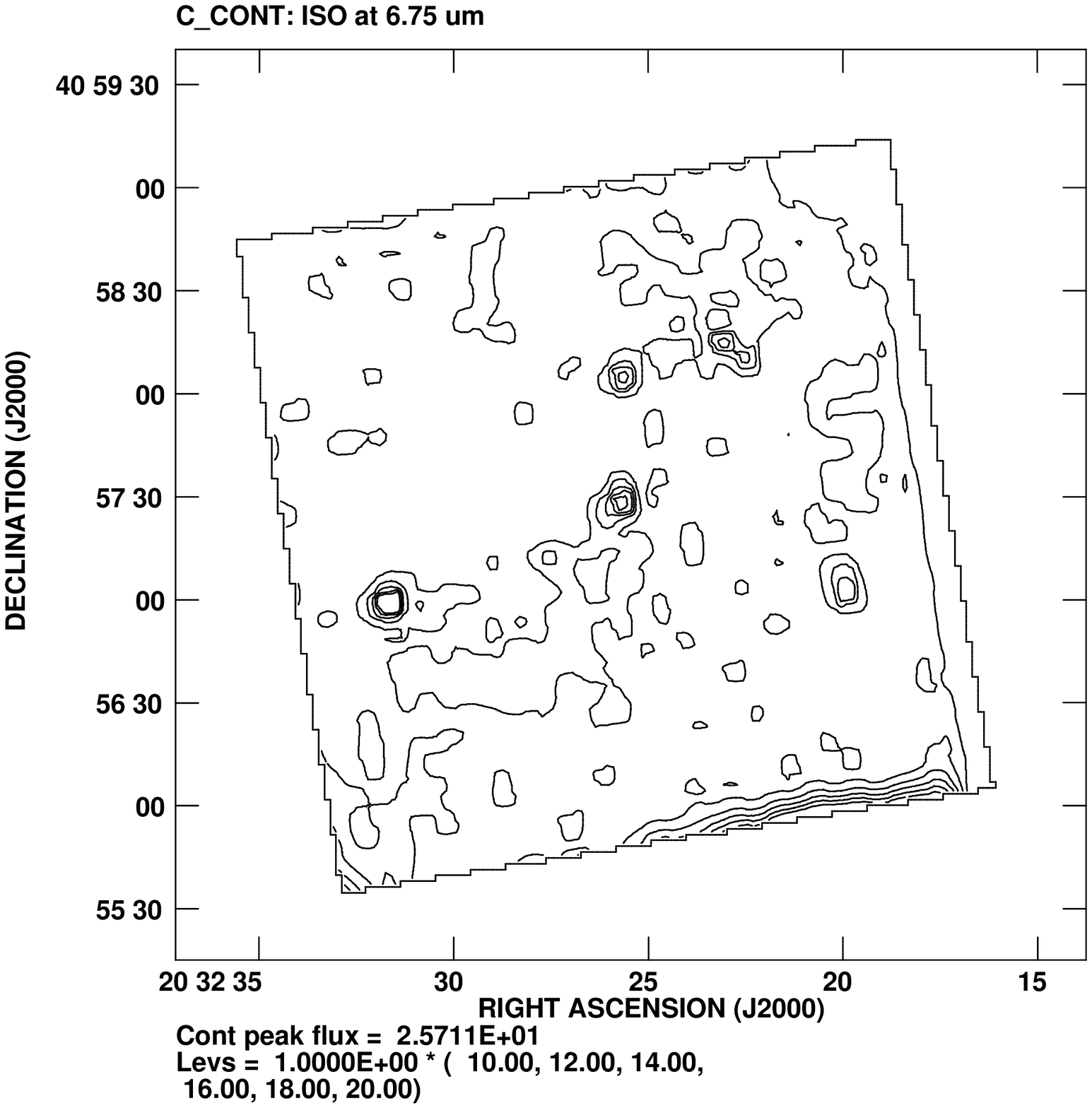,angle=0,width=76mm,clip=}\quad\epsfig{
file=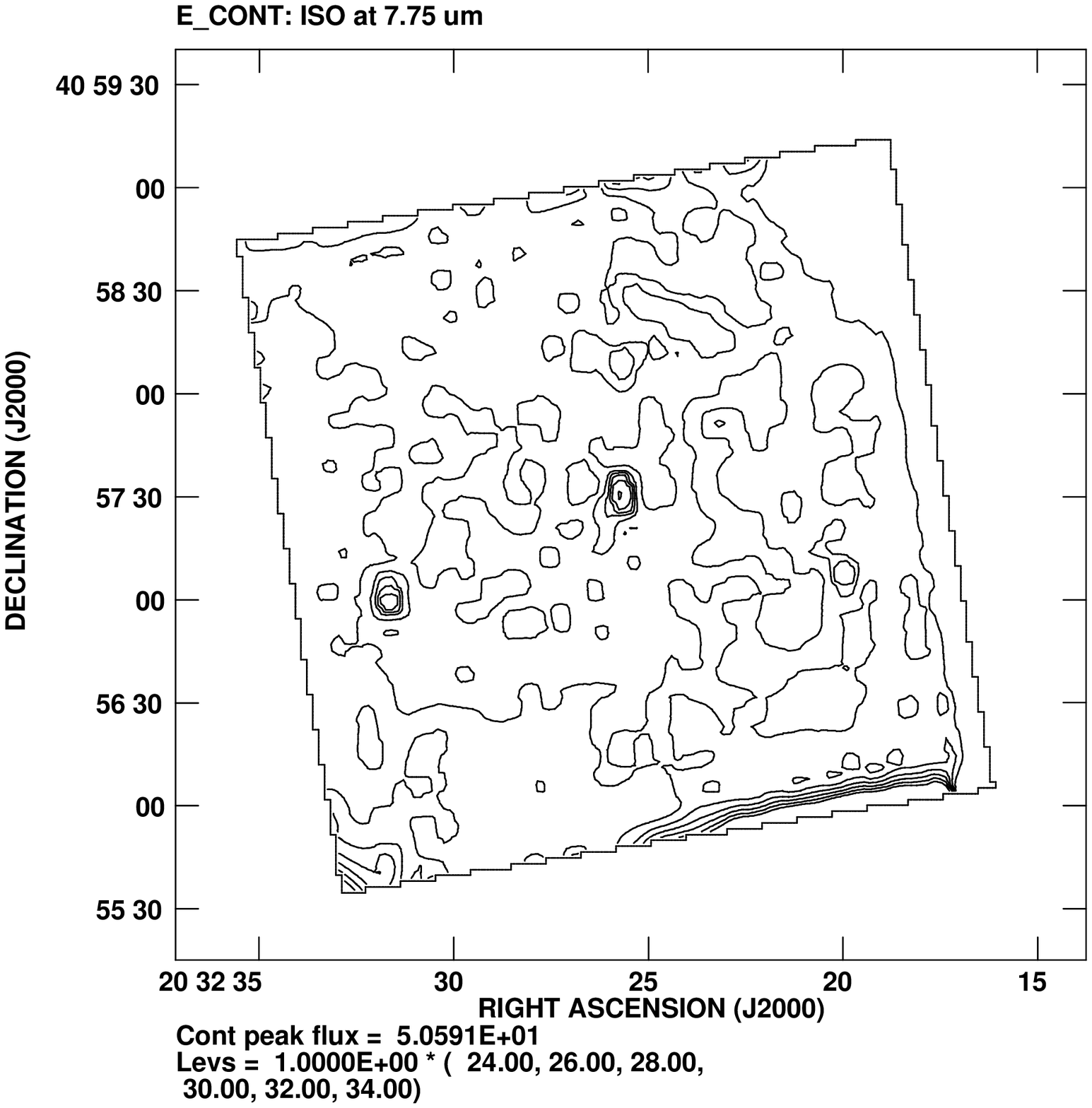,angle=0,width=76mm,clip=}}
\vspace{-1mm}
\centerline{\epsfig{file=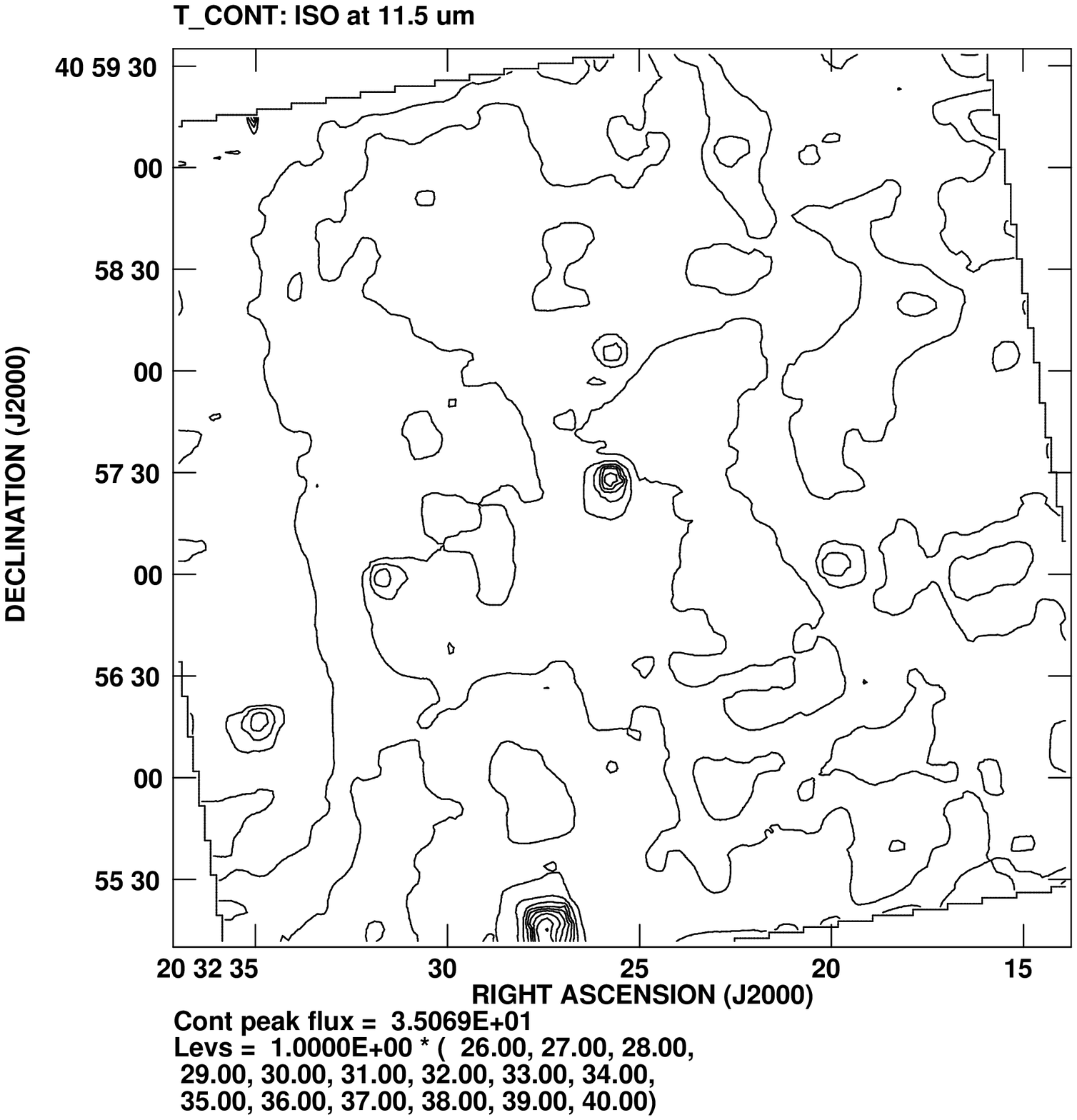,angle=0,width=76mm,clip=}\quad\epsfig{
file=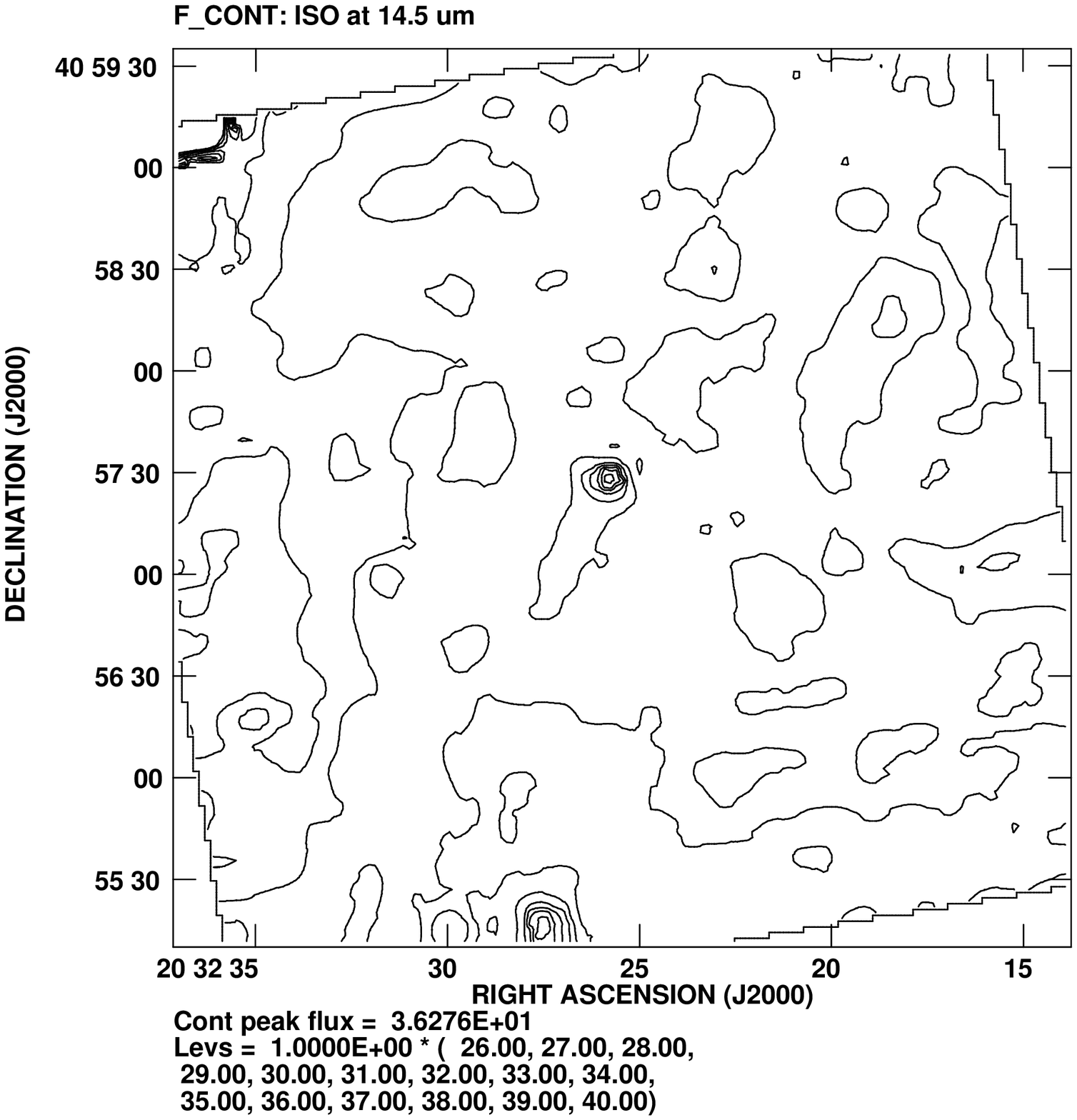,angle=0,width=76mm,clip=}}
\vspace{-1mm}
\caption{ISOCAM images of the Cyg X-3 field in quiescence. Images from
top-left to bottom-right are at 4.5, 6.0, 6.75, 7.75, 11.5 and 14.5
$\mu$m.  Photometry results are presented in Table \ref{fluxes}.}
\label{cam_images}
\end{figure*}

In contrast to the DSS image, Fig.\ \ref{cam_images} shows contour
images at the wavelengths 4.5--14.5 $\mu$m.  The images have been
flat-fielded and dark-subtracted using the techniques described above.
The data were exported and imaged using the NRAO Astronomical Image
Processing System (AIPS), and a rotation of $+99\fdg 17$ was applied
to each image.

The absolute position of the ISOCAM images is unknown due to
mis-alignment after a filter change.  The co-ordinates of a source can
be offset from the true position by as much as 2 pixels, or for our
observations, 12 arcsec.  Positions can be calculated from
phase-referenced radio images, optical sources or catalogued sources.
The positions were calculated from a radio observation of Cyg X-3
using MERLIN at 22 GHz with a 30 mas beam size (Ogley et al.\ 2000).
The radio position has a Gaussian-fitted position of ${\rm
RA}(2000)=20^{\rm h}~32^{\rm m}~25\fs 712$, ${\rm
Dec.}~(2000)=+40^{\circ}~57'~28\farcs 21$, and the {\it ISO} positions
of all sources are correlated to this.

In some images (6.0, 6.75 and 7.75 $\mu$m in Fig.\ \ref{cam_images})
there is a significant misalignment between the flat-field frame and
the array which produces a gradient at the edge of the image.  Other
techniques to flat field the data produced larger errors.  As none of
the sources of interest are near to the affected region, the
photometry presented below is still valid.

\subsection{Photometry}

\subsubsection{Quiescence}

\begin{figure*}
\epsfig{file=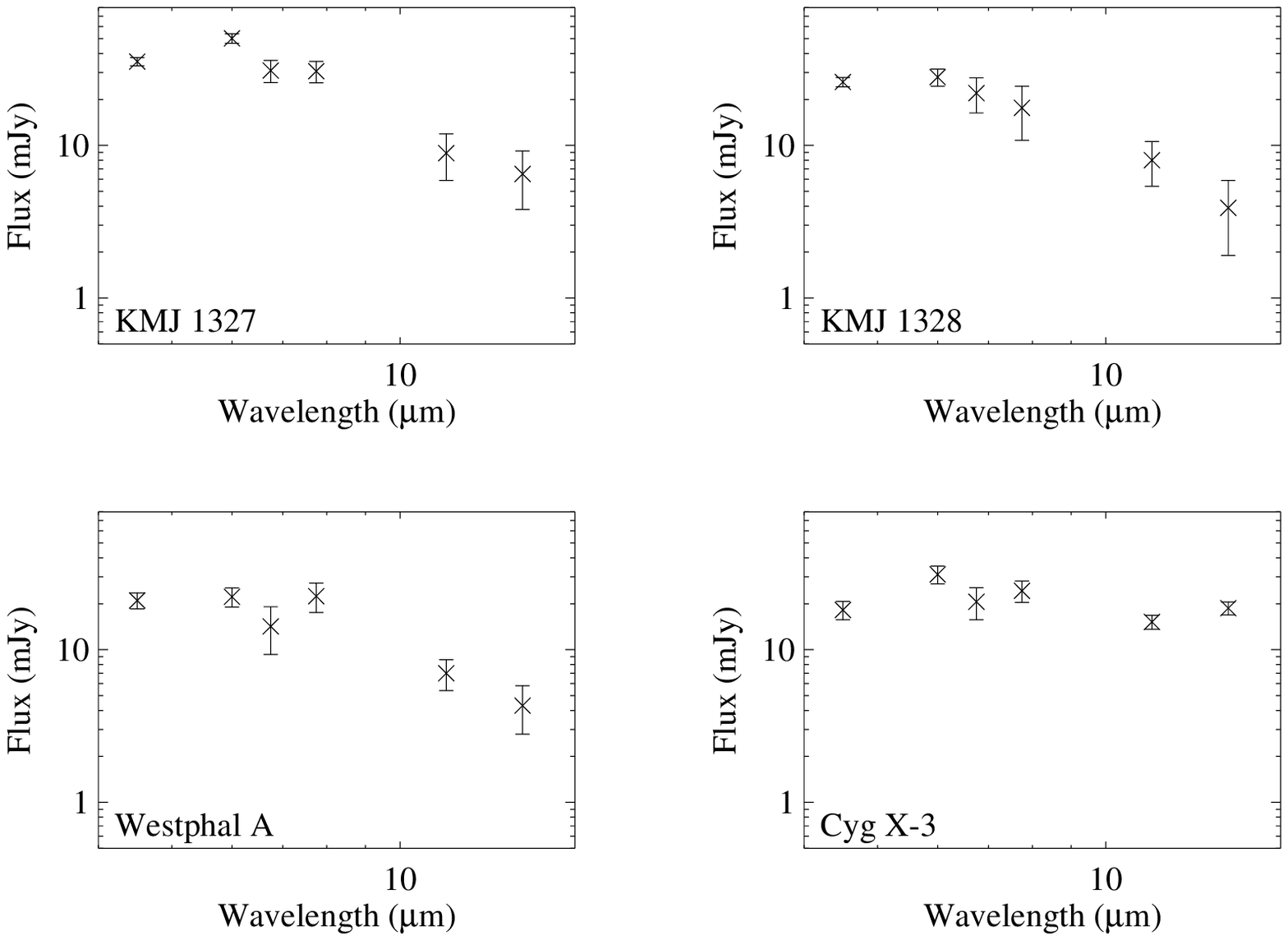, width=6in}
\caption{Spectrum of the four sources, KMJ 1328, 1327, Westphal et
al. `A' and Cyg X-3.  No data have been de-reddened.}
\label{four_star_spectra}
\end{figure*}

Photometry on all four sources was performed using private code by Dr
Siebenmorgen at Vilspa.  An aperture of 12 arcsec was used on all
sources and the flux densities are given in Table \ref{fluxes}.
Spectra of the four sources are shown in Fig.\
\ref{four_star_spectra}.  No data have been de-reddened.


\subsubsection{ISOCAM--outburst}

Photometry of the sources in the three flaring epochs were obtained in
the same way, as in the quiescent observations, with a 12 arcsec
aperture.  The results for the photometry is presented in Table
\ref{four_star_photometry}, for all sources.  Cygnus X-3 is clearly
variable compared to the field stars.

\begin{table}
\caption{Infrared fluxes for the four sources against epoch during a Cyg
X-3 flaring state.  No source has been de-reddened.  The top panel shows
data at 4.5 $\mu$m and the bottom panel shows data at 11.5 $\mu$m.  The
orbital phase of Cyg X-3 is shown using the ephemeris by Matz (1997).}
\begin{tabular}{lccc}\hline
	& \multicolumn{3}{c}{4.5 $\mu$m flux (mJy)} \\
MJD		& 50613	& 50616	& 50619			\\
Date 1997 Jun	& 15	& 18	& 21				\\ 
Cyg X-3 phase	& 0.85		& 0.64		& 0.54 \\
\hline
Cyg X-3		& 28.1 $\pm$ 2.3 & 53.8 $\pm$ 3.1 & 68.5 $\pm$ 3.2 \\
KMJ 1327	& 27.7 $\pm$ 1.9 & 32.9 $\pm$ 2.4 & 32.6 $\pm$ 2.5 \\
KMJ 1328	& 23.9 $\pm$ 2.2 & 24.3 $\pm$ 2.0 & 23.1 $\pm$ 1.9 \\
Westphal et al.\ `A'  & 20.5 $\pm$ 2.4 & 19.5 $\pm$ 2.3 & 19.5 $\pm$ 2.6 \\
\hline
	& \multicolumn{3}{c}{11.5 $\mu$m flux (mJy)} \\
Cyg X-3		& 10.6 $\pm$ 0.6 & 78.2 $\pm$ 0.8 & 40.4 $\pm$ 0.8 \\
\hline
\end{tabular}
\label{four_star_photometry}
\end{table}

\subsection{Spectra}

The Cyg X-3 data have been de-reddened using an extinction coefficient
in the {\it J} band (1.25-$\mu$m) of $A_{J} = 6$ (Fender et al.\
1996), and an extinction law of $\lambda^{-1.7}$ (Matthis 1990).

\subsubsection{Quiescent spectrum}

\begin{figure*}
\epsfig{file=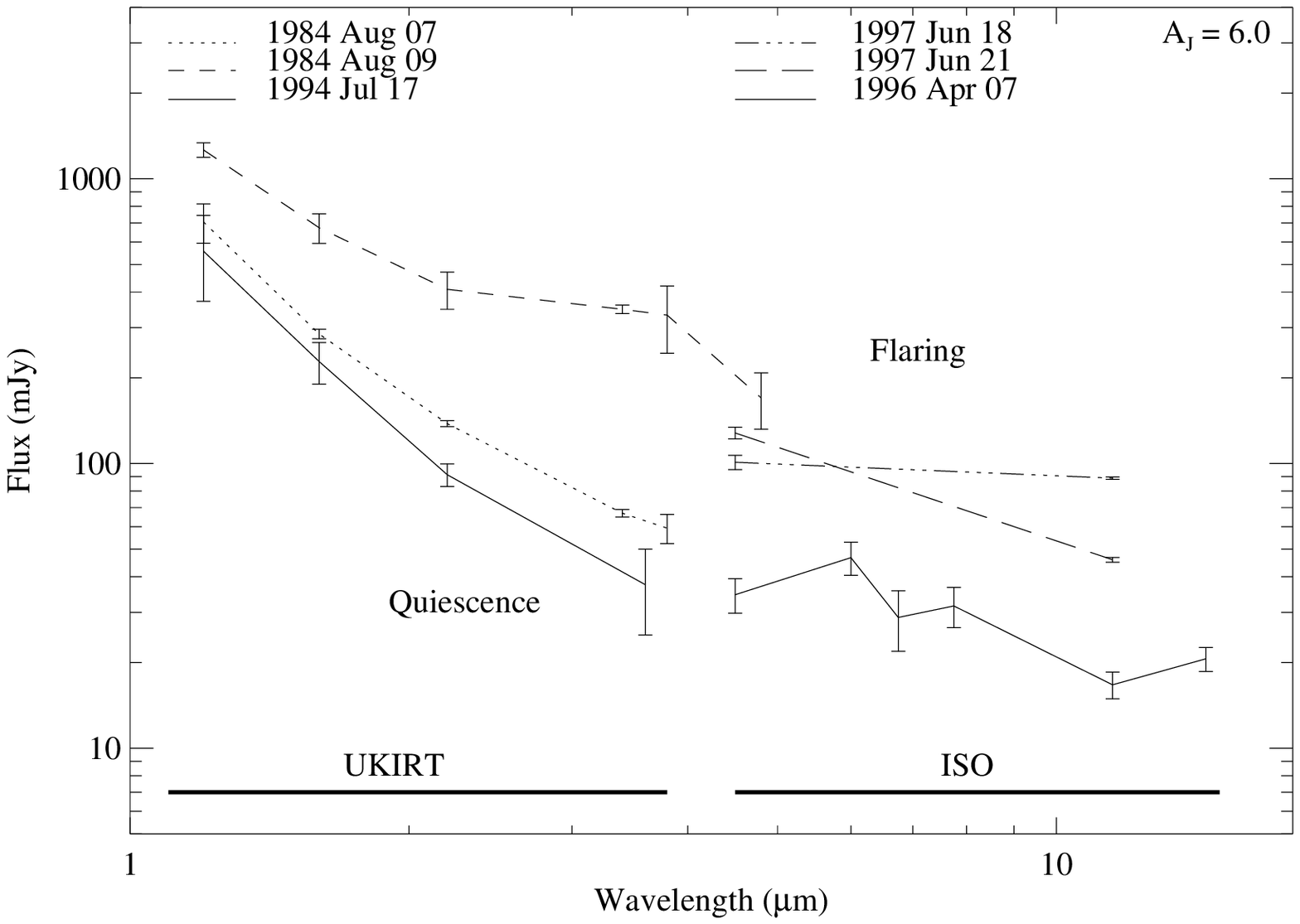, width=6in}
\caption{Quiescent and flaring spectra for Cyg X-3, observed by UKIRT
at 1-4 $\mu$m and {\it ISO} at 4-15 $\mu$m.  Included is UKIRT
photometry from 1984 (Fender et al.\ 1996), and from 1994 at orbital
minimum, from Fender et al.\ (2000).  All data have
been de-reddened using an absorption co-efficient of $A_{J} = 6$
(Fender et al.\ 1996) and a $\lambda^{-1.7}$ dependence (Matthis
1990).}
\label{ukirt_iso_q+f}
\end{figure*}

The de-reddened data for Cyg X-3 in quiescence are plotted in Fig.\
\ref{ukirt_iso_q+f} along with UKIRT data from 1994 July 17 taken at
orbital minimum (Fender et al.\ 1996; Fender, Bell Burnell \& Pooley
2000).  During the 1994 epoch observations a number of short-timescale
flares occurred, but we have been careful only to use data at a time
when the source was in quiescence.  The spectrum shows a flattening at
the longer wavelengths, with a spectral index of 3.0 between 1.25-2.2
$\mu$m and 0.4 between 4.5-14.5 $\mu$m.

\subsubsection{Flaring spectra}

\begin{figure*}
\epsfig{file=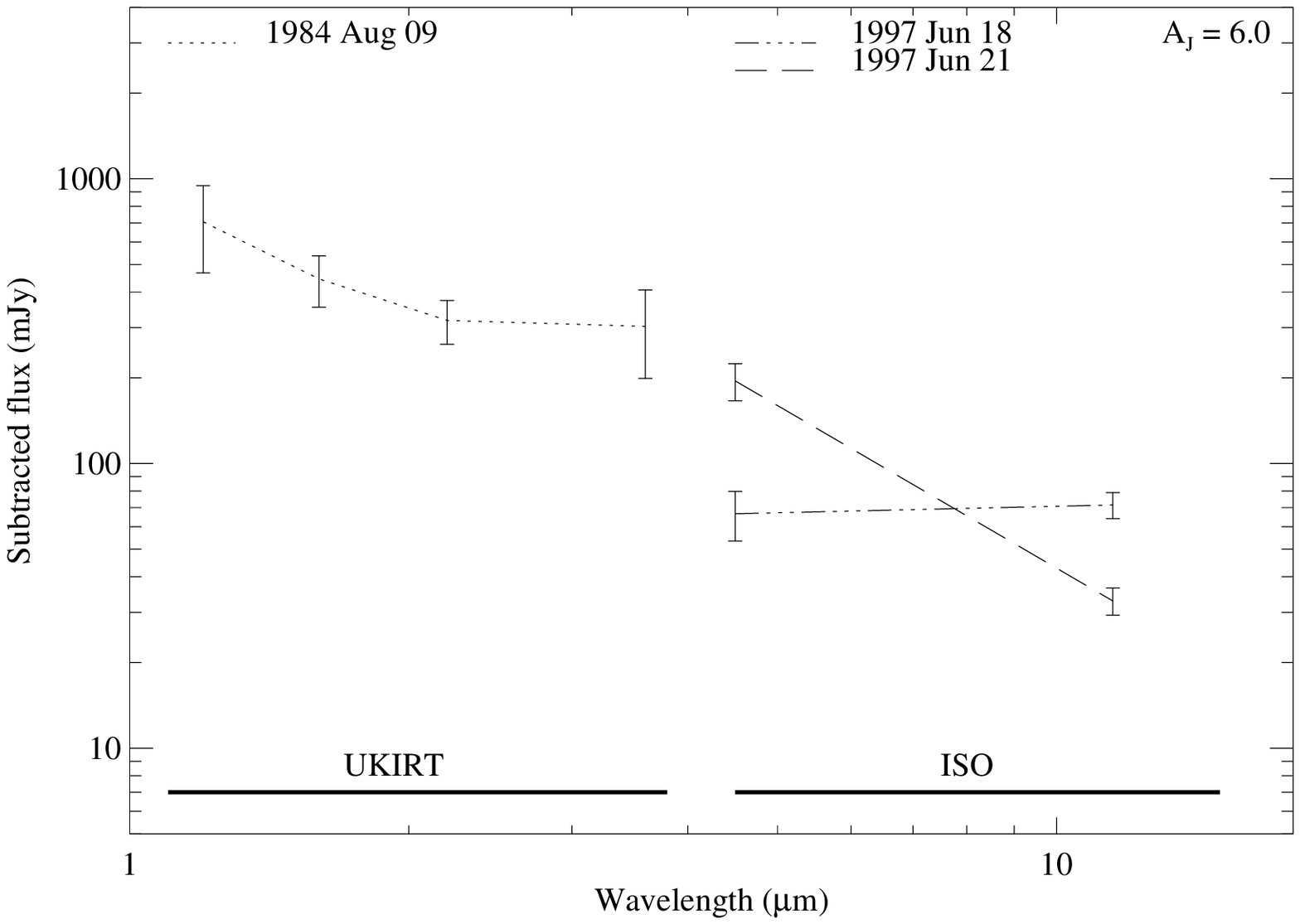, width=6in}
\caption{The flare spectrum with the quiescent spectrum (as shown in
Fig.\ \ref{ukirt_iso_q+f}) subtracted.}
\label{ukirt_iso_f}
\end{figure*}

Table \ref{flare_photometry} shows the de-reddened photometry and
spectral indices for the three observations.  These spectra are also
plotted in Fig.\ \ref{ukirt_iso_q+f}.  The UKIRT data from 1984 is from
Fender et al.\ (1996) and is shown for two nights on 1984 Aug 07, 09.
Figure \ref{ukirt_iso_f} shows the flare spectrum with the quiescent
spectrum subtracted.
\begin{table}
\caption{De-reddened fluxes for Cyg X-3 over the three flaring
epochs.  Data have been de-reddened using an extinction of $A_{J} = 6$
(Fender et al.\ 1996), an extinction law which follows
$\lambda^{-1.7}$ (Mathis 1990) and the spectral index uses the
definition $S \propto \nu^{\alpha}$.}
\begin{tabular}{lccc}\hline
Epoch	& 4.5 $\mu$m flux (mJy) & 11.5 $\mu$m flux (mJy) & $\alpha$ 	\\
\hline
50613	& 52.6 $\pm$ 4.3	& 12.0 $\pm$ 0.7	& 1.6	\\
50616	& 101  $\pm$ 6  	& 87.8 $\pm$ 0.9	& 0.1	\\
50619	& 128  $\pm$ 6		& 45.9 $\pm$ 0.9	& 1.1	\\
\hline
\end{tabular}
\label{flare_photometry}
\end{table}

\section{Discussion}

\subsection{The mass-loss rate} 

In the {\it ISO} regime the spectrum becomes like that predicted by
Wright \& Barlow (1975; hereafter WB75) for a non-accelerated,
spherical wind.  At long infrared wavelengths, WB75 showed that
free-free emission in a wind would display an $S \propto \nu^{2/3}$
spectrum, while at longer radio wavelengths this is modified to
produce a $S \propto \nu^{0.6}$ spectrum.  Our de-reddened quiescent ISO
fluxes can be fitted with a function $S(\nu) = S_{0}
\left(\nu/10^{9}\right)^{\alpha}$ with the parameters $S_{0} = 0.04
\pm 0.06$ and $\alpha = 0.61 \pm 0.13$ to give a reduced $\chi^{2}$
value of 1.74.  The $S_{0}$ parameter is the flux at 1 GHz and its
value around 0 shows there is negligible expected free-free emission
at radio frequencies.

There is no evidence for infrared emission from dust, consistent with
the absence of dust in WNs demonstrated by Cohen (1995).  This means
that the dust responsible for the soft X-ray halo emission reported by
Predehl \& Schmitt (1995), at an angular separation of at least 100
arcsec, cannot lie close to Cyg X-3.

Using the WB75 model to obtain first-order wind parameters, the free-free
emission flux is given by
\begin{equation} 
S = \frac{2.3\times 10^{10}}{D_{\rm kpc}^{2}} \left(\frac{\dot{M}}{\mu
v_{\infty}}\right)^{4/3} \left(\nu_{\rm GHz}\gamma_{\rm
e}gZ^{2}\right)^{2/3}\;\;{\rm mJy},
\end{equation} 
where $D_{\rm kpc}$ is the distance to the source in kpc, $\dot{M}$ is the
mass lost in the wind in $\rm M_{\odot}$ yr$^{-1}$, $\mu$ is the mean
atomic weight per nucleon, $v_{\infty}$ is the velocity of the wind at
infinity measured in km s$^{-1}$, $\nu_{\rm GHz}$ is the observing
frequency in GHz, $\gamma_{\rm e}$ is the number of free electrons per
nucleon, $g$ is the Gaunt factor and $Z$ is the mean ionic charge. From
our {\it ISO} data, the quiescent flux at a wavelength of 6.75$\mu$m
(44,400 GHz) is $30 \pm 10$ mJy. Substituting these values into equation
(1) together with a distance of 10 kpc, and assuming a value for the Gaunt
factor of unity, gives an expression for the mass-loss rate of
\begin{equation} 
\dot{M} = 3.2 \times 10^{-8}\; \mu v_{\infty} \left(\gamma_{\rm e}
Z^{2}\right)^{-1/2}\;\;{\rm M_{\odot}\;yr^{-1}}.
\end{equation} 
The velocity of the wind in Cyg X-3 has been investigated by a number of
authors.  van Kerkwijk et al.\ (1992; 1996) observed Cyg X-3 using a high
resolution spectrograph on UKIRT. Their observations were scheduled over a
range of orbital phases and from this they were able to obtain a wind
velocity of $v_{\infty} = 1600 \pm 300$ km s$^{-1}$.  More recently, the
flattened wind model of FHP99 gives an outflow velocity of $v_{\infty}
\sim 1500$ km s$^{-1}$. It is this latter value that we will adopt in this
paper.

The other three variables which are unknown -- $\mu, \gamma_{\rm e}$ and
$Z$ -- are all related to the wind's composition.

(a) {\it A WN-type wind}.  Using a wind composition typical of late WR
stars, $\mu=1.5$, $\gamma_{\rm e}=1$, $Z=1$ (Waltman et al.\ 1996) and the
flux at 6.75$\mu$m (44,400 GHz) the mass-loss rate is $\dot{M}_{\rm a} =
7.8 \times 10^{-5}\;\rm M_{\odot}$ yr$^{-1}$.

(b) {\it A partially ionised wind}. The composition of this wind was
used by Waltman et al.\ (1996) to explain the time difference between
radio flares.  This wind has the parameters of $\mu=4$, $\gamma_{\rm
e}=0.5$, $Z= 0.5$.  The mass-loss rate using our infrared fluxes and
these values is $\dot{M}_{\rm b} = 2.9 \times 10^{-4}\;\rm M_{\odot}$
yr$^{-1}$.

(c) {\it A fully ionised wind}.  van Kerkwijk et al.\ (1996) used this
type of wind when interpreting UKIRT data for the observations of the
Wolf--Rayet-like lines.  This wind has the parameters of $\mu=4$,
$\gamma_{\rm e}=2$, $Z= 2$ and using our infrared fluxes a mass-loss
rate of $\dot{M}_{\rm c} = 3.6 \times 10^{-5}\;\rm M_{\odot}$
yr$^{-1}$ is calculated.  

We note that Crowther et al.\ (1995) have found averages of
$\mu=3.23$, $\gamma_{\rm e}=1.010$, $Z=1.015$ for Galactic WN(L) stars
(WN7-8) which would infer a mass-loss rate of $\dot{M} = 8.1 \times
10^{-5}\;\rm M_{\odot}$ yr$^{-1}$.  For a WN(E) star (WN4-6) one might
expect $\mu \simeq 4$, $Z = 1$, $\gamma = 1$ (e.g.\ Willis 1991)
giving $\dot{M} \simeq 2.1 \times 10^{-4}\;\rm M_{\odot}$ yr$^{-1}$.

Thus we find that for a single temperature, non-accelerating, spherical
wind and for a variety of ionisation states, the free-free emission
causing the infrared fluxes in the {\it ISO} wavelengths is due to a wind
with a mass loss rate of (0.4--2.9)$\times 10^{-4}\;\rm M_{\odot}$
yr$^{-1}$.

This value for the mass loss rate is towards the upper end of the
range observed for (single) WR stars, and is slightly higher than that
predicted by Waltman et al.\ (1996) to be responsible for the delay
between radio emission at a range of frequencies.  However, it is
slightly lower than is predicted by van Kerkwijk et al.\ (1996) for
their interpretation of the short wavelength infrared fluxes.  Results
from both papers together with our fluxes are given in Table
\ref{massloss_rates}.  

More significantly, there is a discrepancy in the value of $\dot{M}$
derived from the orbital period change and the value from our {\it
ISO} results.  Using the relationship $\dot{P}/2P = \dot{M}/M_{T}$ and a value for $\dot{P}$ of $-6.6 \times 10^{-10}$ s s$^{-1}$, Kitamoto et al.\ (1995) calculated a value of $\dot{M}
= 0.6 \times 10^{-6}\;M_{T}\;\rm M_{\odot}\;yr^{-1}$ where $M_{T}$ is
the total system mass assumed to be of the order $10\;\rm M_{\odot}$.
This is an order of magnitude lower than that from our {\it ISO}
results $(10^{-4}\;\rm M_{\odot}\;yr^{-1})$.  This difference between
$\dot{M}$ derived from the orbital period change, and $\dot{M}$
derived from thermal emission observations (with the latter being the
larger) is not unique to Cyg X-3.  The mass-loss rate for the binary
V444 Cyg (WR139) has been calculated to be $6 \times 10^{-6}$ and $2.4
\times 10^{-5}$ M$_{\odot}$ yr$^{-1}$ by the $\dot{P}_{\rm orb}$
method and thermal emission method respectively (Underhill et
al. 1990; Prinja et al. 1990).

\begin{table}
\caption[Mass loss rates for various wind models]{The mass loss rate
for the Wolf--Rayet star given by the various models discussed in the
text.}
\begin{minipage}{3in}
\begin{tabular}{lccc}
\hline
Method	& $\dot{M}_{\rm a}$  & $\dot{M}_{\rm b}$  & $\dot{M}_{\rm c}$\\
			& \multicolumn{3}{c}{$\times 10^{-5}\;
			\rm M_{\odot}$ yr$^{-1}$} \\
\hline
Radio photospheres$^{1}$\footnotetext{$^{1}$ Waltman et al.\ (1996)}	
	& $\leq 0.25$ & $\leq 1.65$	& $\leq 2.7$ 	\\
This paper		
	& $7.8 \pm 2.1$ & $29 \pm 8$	& $3.6 \pm 1.0$	\\
Near-IR fluxes$^{2}$\footnotetext{$^{2}$ van Kerkwijk et al.\ (1996)}
	& --	& --	& $\geq 12$	\\
\hline
Orbital period change$^{3}$\footnotetext{$^{3}$ Kitamoto et al.\ (1995)}	& 0.06 	& -- & -- \\
\hline
\end{tabular}
\end{minipage}
\label{massloss_rates}
\end{table}

\subsubsection{$\dot{M}$ from $S$ - an examination of assumptions}

In obtaining $\dot{M}$ from the {\it ISO} data we have assumed that
the wind is isothermal, at a constant velocity, shows spherical
symmetry, stationarity and homogeneity. We have further assumed that
the free-free flux is not seriously contaminated by stars Z and D
identified on the Cyg X-3 finder chart and within the {\it ISO} pixel
size (Fender \& Bell Burnell 1996) nor by synchrotron emission from
the tail of the radio-mm spectrum. The agreement with the Wright and
Barlow model suggests that these assumptions (with the possible
exception of spherical symmetry) are reasonable. We have assumed a
distance of 10 kpc to the source; since $\dot{M}$ scales as $D^{3/2}$,
to effect a 10-fold reduction in the mass-loss rate, the distance
would have to reduced to $\sim$ 2 kpc.  The extinction to Cyg X-3 is
uncertain with $4.5 < A_{J} < 7.5$.  This corresponds to range of
extinction at 6.75 $\mu$m of 0.26-0.43, and an uncertainty of only 10
per cent in the value of the flux.

We have also assumed that the Gaunt factor, $g$, is unity.  In fact it
depends on $Z$, $T$ and the observing frequency, and occurs in the
expression for $\dot{M}$ as $g^{-1/2}$.  The equation for the Gaunt
factor given by Spitzer (1962), and subsequently used by Leitherer \&
Robert (1991) is valid for $\nu_{p} << \nu << kT/h$, where $\nu_{p}$
is the plasma frequency.  This condition is satisfied over the range
of parameters we use here, and taking extreme values for the temperature
of 10,000 to 350,000 K.  Spitzer's equation gives values of $g$
ranging from 3.4 to effectively zero (at the lowest temperature and
highest $Z$).  Alternatively Sutherland (1998) determines the Gaunt
factor for the spectral range submillimetre to hard X-ray and for
temperatures from 10 to $10^9$ K.  We believe that in his table 2 the
labelling of the rows and columns has been interchanged, and that
figures 2b and 2c have been switched.  Sutherland's work then gives
Gaunt factors that range from 1.5 to 3.2.  Sutherland considers
photoionisation (which may be more appropriate to Cyg X-3) as well as
collisional ionisation and shows that the photoionisation figures for
the Gaunt factor are typically a factor of 2 lower.  Our assumption
that the Gaunt factor is unity well represents the middle of the range
obtained, but it is uncertain by a factor of $\sim 3$.

Nugis, Crowther \& Willis (1998) model the mass-loss rate in a wind
that is clumped, and in which both the ionisation structure and the
clumping changes with radial distance.  Selecting values for their
free parameters that best fit for each star the flattening spectral
slope from infrared to radio, they obtain mass-loss rates for clumped
winds for 15 WN stars.  There is considerable scatter in their
results, but they obtain an average of $\log{\dot{M}(\rm clumpy)} -
\log{\dot{M}(\rm smooth)} = -0.2 \pm 0.3 \; (1 \sigma)$.  In other
words, their clumpy wind model on average reduces the value of
$\dot{M}$ to $0.6^{+0.6}_{-0.3}$ of the $\dot{M}(\rm smooth)$ value.
Assuming Cygnus X-3 is close to average, then again this modification
alone is insufficient to explain the discrepancy.

FHP99 show that the most likely interpretation of their infrared
spectra and their observation of an orbital modulation of \textit{V/R}
involves a non-spherical wind, with a flattened, disc-like enhanced
wind, probably in the plane of the binary. They note that this also
reconciles the high infrared emissivity with an optical depth to the
X-ray source that remains small (as long as the system is not viewed
edge-on). Adopting their model of an inclined disc of enhanced wind,
the Gorenstein (1975) approximation suggests the maximum X-ray column
observed is consistent with a maximum value of $\dot{M} \sim 10^{-5}$
$\rm M_{\odot}$ yr$^{-1}$ along the line of sight to the X-ray
source. If the disc-wind has a solid angle of 10 per cent of $4\pi$
then, to be consistent with the {\it ISO} data and maximum column, a
mass-loss rate of $1 \times 10^{-3}$ $\rm M_{\odot}$ yr$^{-1}$
sterad$^{-1}$ in the disc-wind is required; for 25 per cent of $4\pi$
it is $4 \times 10^{-4}$ $\rm M_{\odot}$ yr$^{-1}$ sterad$^{-1}$; and
for 50 per cent of $4\pi$ it is $2 \times 10^{-4}$ $\rm M_{\odot}$
yr$^{-1}$ sterad$^{-1}$ with $10^{-5}\;\rm M_{\odot}\;yr^{-1}$ outside
the disc; i.e. disc density enhancements of 100, 40 and 20
respectively.

We compare these results with previously reported theoretical and
observational work on deviations from spherical symmetry and density
enhancements.  The pioneering work of Schmid-Burgk (1982) demonstrated
that deviations from spherical symmetry did not change the spectral
slope but led to over-estimates of the
mass loss rate.  However, the maximum correction factor found was only
a factor of 2, for a pancake (disc) with a 10 to 1 aspect ratio.
Ignace, Cassinelli \& Bjorkman (1996) investigated density
enhancements caused by rotation.  For a WN5 wind their model found a
density approximately 3 times greater than the global average in an
equatorial cone of 10 degrees half-angle (or pole-to-equator density
ratio of about 6).  Rotation at only 10 per cent of critical was
required.

Harries, Hillier \& Howarth (1998) found that 15-20 per cent of WR
stars have significant polarization, which they attribute to global
wind asymmetries, found only in the fastest rotators.  They also found
that these stars have the highest values of $\dot{M}$.  They postulate
that these stars have equator to pole density ratios of 2 or 3 (with
$\dot{M}$ scaling linearly with density).  St Louis et al.\ (1995)
propose a sectored wind structure for the WN star WR6.  They find the
wind divides roughly 50:50 into fast and slow wind regions, with the
higher continuum fluxes coming from the hotter, higher speed regions.
Radio imaging of the WN8 star WR147 (Williams et al., 1997) report a
wind `shape' with a 3 to 2 aspect ratio.  There have been suggestions
of a disc in the WR6 star WR134 - see Vreux et al.\ (1992), and
references therein.  Nota et al.\ (1995) report a bipolar outflow in
the nebula around the WN8 star WR124, with an opening angle between 30
and 45 degrees.  The binary nature of Cyg X-3 is likely to create
deviations from spherical symmetry, as proposed originally by van
Kerkwijk (1993) and for WR140 by Williams et al.\ (1990).  The disc
configuration reported by White \& Becker (1995) is disputed by
Harries et al.\ (1998).

If the WR star's rotation is tidally locked to the 4.8 h orbital
period, then it will be rotating at close to its critical velocity and
an equatorial enhancement of the mass loss rate is likely.  We
conclude that deviations of the wind from spherical symmetry need to
be taken into consideration. Ignoring such asymmetry tends to
overestimate the the mass-loss rate. However, to be consistent with
the visibility of the X-ray source the density contrast required by
our result appears to be large, and we are not aware of any work
quantifying the overestimate for such large disc density enhancements.
 
\subsubsection{$\dot{M}$ from $\dot{P}$ - an examination of assumptions}

To obtain the total stellar mass-loss rate which drives the orbital
changes any other mass-loss rates occurring need to be added to this
value of wind mass-loss rate. The X-ray luminosity of Cyg X-3 implies
an accretion rate on to the compact object of $\sim 10^{-8}$ $\rm
M_{\odot}$ yr$^{-1}$, which is small compared with the mass lost
through the wind.  Mass will also be lost into the radio jets, but this
is an uncertain quantity, and the mass in the jets could
be anything between 0.1 to 100 per cent of the in-flowing mass (Fender
\& Pooley 1998, 2000).  These additional masses are in the sense to
exacerbate the problem.

In obtaining $\dot{M}$ from $\dot{P}_{\rm orb}$ several other
assumptions are made: that there is no loss of angular momentum other
than the specific orbital angular momentum of the WR star carried away
by the wind; that the rotating bodies are rigid; that the 4.8-h
modulation, whatever it is, is rigidly locked to the orbital period;
that the orbital period is 4.8-h, and not some multiple of that
figure; that the eccentricity is zero; and that there are no other
dynamical effects (e.g. gravitational radiation) or torques
(e.g. magnetic) affecting the value of $\dot{P}_{\rm orb}$.

The assumption that the eccentricity is small seems justified - given
the amount of circumstellar matter a circular orbit seems likely, and
no change in the light curve which could be attributed to precession
has been observed.  For convenience we adopt $e=0$.  It seems likely
that 4.8 h is the fundamental period, but until its nature is
understood there remains the possibility that the 4.8 h modulation is
not rigidly tied to the orbital period and that the $\dot{P}$ observed
is not accurately that of the binary system.

Additional angular momentum could be lost through the jets, or through
magnetic torques, for example, and ignoring such losses would cause
$\dot{P}_{\rm orb}$ to be underestimated.

Gravitation radiation results in the orbital size shrinking, so
neglect of this effect will cause the $\dot{P}_{\rm orb}$ due to the
mass loss to be underestimated. The effect is most extreme for the
fastest orbital velocities which occur for the most massive components
(assuming the eccentricity is zero). Taking a Wolf-Rayet mass of 10
$\rm M_{\odot}$ and a black hole mass of 40 $\rm M_{\odot}$,
$\dot{P}_{\rm GR} \sim - 6 \times 10^{-11}$, or about 10 per cent of
the observed $\dot{P}$.  If the orbital eccentricity were 0.5, the
effect would be approximately 5 times larger, so although
gravitational radiation produces an effect with the correct sign, it
is not probably large enough to explain the discrepancy. Orbital
motion in a viscous medium (the wind) will also cause the orbit to
shrink.

It is known that tidal effects in early-type stars give rise to
gravity waves which are radiatively damped, leading to circularisation
of elliptical orbits and the synchronisation of orbital and stellar
rotation rates (Kumar \& Qutaert 1998).  Kumar and Qutaert show that if
the stellar rotation rate is less than the orbital rotation rate
(strictly periastron rate), $\dot{P}_{\rm orb}$ is negative.  If we
assume the Wolf--Rayet star rotates like its predecessor O star, then
its rotational period would be a factor 10-20 times longer than the
(current) orbital period.  Tidal forces will presumably have spun it
up; if it is now (close to) synchronised then for a $3\;{\rm
R_{\odot}}$ star it will have a mean equatorial velocity of $\simeq
750$ km s$^{-1}$ and, assuming a mass of 10 $\rm M_{\odot}$, be
rotating at 95 per cent of its critical velocity.  It seems likely that
there have been tidal forces which have tended to contract the orbit
and diminish the observed $\dot{P}_{\rm orb}$, but in the absence of
information about the current rotational period of the Wolf--Rayet we
are unable to estimate the current magnitude of this effect.

In summary, there are several very probable effects which could cause
$\dot{P}_{\rm orb}$ to be underestimated - namely unrecognised sources of
angular momentum loss, tidal effects, and to a lesser extent energy loss
through viscous drag and gravitational radiation. Of the two methods of
determining the mass loss rate, this seems to have the larger number of
uncertainties.

\subsection{Flare data}

It is clear from the observations in 1997 June (Fig.\
\ref{ukirt_iso_q+f} \& \ref{ukirt_iso_f}) that the ISO data reveal
bright flaring at 4.5 and 11.5 $\mu$m at the same time as enhanced
radio and X-ray activity (while not presented here, the X-ray
monitoring data, which reveal an X-ray bright state at this time, are
presented in FHP99). In addition, the spectrum appears to become
flatter (although we caution that the lack of absolute simultaneity of
the observations in the different filters adds additional
uncertainties to the interpretation of spectral indices). This
flattening is in qualitative agreement with the observations of Fender
et al. (1996, 2000) who find that during both rapid (minutes
timescales) and more gradual (days) variability, Cyg X-3 generally
becomes redder in the infrared as it brightens.

While Fender et al.\ (1996) interpret the flaring component from Cyg
X-3 in terms of thermal free-free emission from a hot ($T \geq 10^6$
K) gas, infrared oscillations from GRS 1915+105 (Fender \& Pooley 1998
and references therein) have been interpreted as being synchrotron in
origin. Either way, the rapid, large-amplitude, broad-band variability
in the infrared fluxes during outburst periods implies the occurrence
of extremely violent processes on relatively small physical
scales. While the timescales do not preclude a simple change in the
brightness of the thermal (wind) emission, the significant spectral
flattening appears to rule this out. Instead, similarly to GRS
1915+105, Cyg X-3 appears to display an approximately flat-spectrum
($\alpha \sim 0$) component during outbursts. This may be related to
the flat spectral component now also known to be present in Cyg X-1
(Fender et al.\ 2000). The nature of this component is uncertain, but
it appears to be related to jet formation, and possibly to
instabilities in the accretion flow/disc, at least in GRS 1915+105. In
summary, while there may be variability in the strength of the
underlying thermal (wind) component, we believe the additional flux
observed in outburst to be dominated by a variable, flat-spectrum
component which is probably associated with the jets.

\section{Conclusions}

We report the first quiescent and flaring 4-16 $\mu$m observations
of Cygnus X-3.

The quiescent spectrum is a continuation of the shorter wavelength IR
spectrum, and fits well the WB75 model for thermal free-free emission
from a wind.  We obtain a mass loss rate, assuming spherical symmetry
of $\sim 10^{-4}$ $\rm M_{\odot}$ yr$^{-1}$, and note that while this
is consistent with other determinations, it is an order of magnitude
larger than both the mass loss rate required to derive the change in
orbital period (assuming $M_{\rm T}$ of the order $10\;\rm
M_{\odot}$), and the mass loss rate expected from the hydrogen column
density to the X-ray source.  These measurements are based on
assumptions, which have been scrutinised carefully; while we can
identify a number of areas where factors of a few can be gained, we
have been unable to identify a single area where a factor of ten can
be found.  We agree with the conclusion of FHP99 that the wind is most
likely spherically asymmetric, but if this is the only effect
operating, large disc density enhancement factors are required.  There are
several phenomena which could cause $P_{\rm orb}$ to be underestimated; we
regard this determination of $\dot{M}$ to be less reliable.

During radio and X-ray outburst, the broadband infrared spectrum shows
rapid variability and a spectral flattening associated with increased
flux densities. This is in agreement with what is seen in the near-IR
during rapid (minutes) flaring events (Fender et al.\ 1996). Similar
(but not identical) near-IR events observed from GRS 1915+105 appear
to be associated with jet formation episodes. The emission mechanism
responsible for this flat spectral component remains uncertain; both
free-free and synchrotron emission have been proposed already in the
literature, and await detailed studies to test their
viability. Whatever the origin of this component, it seems likely that
it is associated with hot gas in the jet from the system, which
naturally explains its emergence during periods of enhanced radio
(jet) activity.

\section{Acknowledgments}

The authors wish to thank staff at the {\it ISO} data centres at the
Rutherford Appleton Laboratory in the UK, Saclay in France and Vilspa
in Spain.  The authors are especially grateful for the dedication of
Joris Blommaert and Ralph Siebenmorgen at Vilspa for their help with
reduction of the data.  SJBB thanks Princeton University for
hospitality while this paper was being written.

These data are based on observations with ISO, an ESA project with
instruments funded by ESA Member States (especially the PI countries:
France, Germany, the Netherlands and the United Kingdom) with the
participation of ISAS and NASA.  The ISOCAM data presented in this
paper was analysed using `CIA', a joint development by the ESA
Astrophysics Division and the ISOCAM Consortium.  The ISOCAM
Consortium was led by the ISOCAM PI, C. Cesarsky, Direction des
Sciences de la Mati\`{e}re, CEA, France.  MERLIN is a national
facility operated by the University of Manchester on behalf of PPARC.
The Green Bank Interferometer is a facility of the National Science
Foundation operated by the NRAO in support of NASA High Energy
Astrophysics programs.

\end{document}